# Компьютерная модель обучения с изменяющимся коэффициентом забывания

Майер Р.В.

*Аннотация*—При компьютерном моделировании процесса обучения обычно предполагается, что все элементы учебного материала забываются с одинаковой скоростью. Но на практике те знания, которые включены в учебную деятельность ученика, запоминаются значительно прочнее и забываются медленнее, чем знания, которые он не использует. С целью более точного исследования дидактических систем предложена модель обучения, учитывающая, что при увеличении числа обращений ученика к данному элементу учебного материала: 1) время его использования уменьшается, стремясь к некоторому пределу; 2) коэффициент забывания уменьшается, стремясь к нулю. Рассмотрена компьютерная модель, представлены программы на языке Pascal, приведены результаты моделирования, проведен их анализ.

*Ключевые слова*— дидактика, информационно-кибернетический подход, компьютерное моделирование процесса обучения.

## I. Введение

Одна из актуальных проблем теории обучения состоит в построении математической и компьютерной модели дидактического процесса [1-9]. Последнее предполагает создание компьютерной программы, которая ведет себя как система "учитель-ученик", и проведение серии вычислительных экспериментов при различных параметрах, начальных условиях и внешних воздействиях. Высокое быстродействие современных ЭВМ позволяют обрабатывать большие объемы информации и достаточно быстро осуществлять компьютерную имитацию. Изменяя начальные данные и параметры модели, можно исследовать пути развития системы, определить ее состояние в конце обучения. В этом состоит преимущество данного подхода по сравнению с методом качественного анализа. Логичность и формализованность, воспроизводимость и конкретность получающихся выводов выгодно отличают метод имитационного моделирования от "метода качественных рассуждений". При этом используются дискретные и непрерывные модели [1, 2, 4, 8, 9], метод статистических испытаний, а также мультиагентное моделирование [3], при котором каждый учащийся заменяется программным агентом, функционирующим независимо от других агентов.

При моделировании обучения обычно исходят из того, что все элементы учебного материала (ЭУМ) одинаково хорошо усваиваются и одинаково медленно забываются. На самом деле это не так: те ЭУМ, которые включены в деятельность ученика забываются существенно медленнее, чем ЭУМ, которые были один раз изучены и не используются. Можно предположить, что учет этого фактора позволит создать компьютерную модель обучения, которая в большей степени соответствует реальной ситуации.

## II. Обучение в результате многократного повторения одного ЭУМ на уроке

Рассмотрим ученика, который в процессе обучения вынужден решать последовательность однотипных задач по одной и той же теме. Например, он в течение урока должен в определенные моменты времени складывать числа (или читать отдельные слова, выполнять задания теста). Остальное время на уроке он занимается другой учебной деятельностью, которая нас не интересует. Пусть в момент $t_i$ ученик начинает решать задачу в $i$-ый раз, при этом уровень усвоения соответствующего ЭУМ у него увеличивается до $Z=1$. Учтем, что время решения задачи $\tau$ (или время, затрачиваемое на работу с данным ЭУМ), зависит от того, сколько раз $s$ это задача решалась ранее. Можно предположить, что с ростом $s$ время $\tau$ уменьшается по закону: $\tau = 1+1{,}5e^{-s/5}$ условных единиц времени (УЕВ), стремясь к $\tau_\infty = 1$ УЕВ. Выполнив задание и повысив уровень знания соответствующего ЭУМ до 1, ученик переключается на решение другой учебной задачи и начинает забывать усвоенный ЭУМ в соответствии с законом забывания $dZ/dt = -\gamma \cdot Z$. Будем исходить из того, что при увеличении числа $s$ использования данного ЭУМ он запоминается лучше, то есть коэффициент забывания уменьшается, например, по такому закону: $\gamma = 0{,}002e^{-s}$ (УЕВ$^{-1}$). Ниже приведен текст компьютерной программы ПР–1 в среде Free Pascal, которая моделирует обучение при использовании ЭУМ в моменты времени 3, 6, 9, 12, 15, 18 УЕВ.

ПР–1.

```
Program Izuchen_1_EUM;
{$N+}Uses crt, graph;
```

Майер Роберт Валерьевич, доктор педагогических наук, профессор кафедры физики и дидактики физики ГОУ ВПО "Глазовский государственный педагогический институт" (email: robert_maier@mail.ru)

```
Const N=6; dt=0.001; Mt=20;
t1: array[1..6] of single=(3,6,9,12,15,18);
Var t,Z,g: single; i,s,Gd,Gm: integer;
BEGIN
Gd:= Detect; InitGraph(Gd,Gm,'c:\bp\bgi'); t:=-3;
Repeat t:=t+dt;
For i:=1 to N do If (t>t1[i])and(s=i-1) then
 begin s:=i; Z:=1; t:=t+1.5*exp(-s/5); end;
 g:=0.002*exp(-s/1); Z:=Z-g*Z;
 circle(10+round(Mt*t),400-round(200*Z),1);
until KeyPressed; CloseGraph;
END.
```

Результаты моделирования представлены на рис. 1. Видно, что после первого и второго обращения к данному ЭУМ приобретенные знания быстро забываются, а после пятого и шестого — забываются очень медленно. То есть в результате многократного использования данного ЭУМ коэффициент забывания уменьшается практически до 0, информация прочно запоминается.

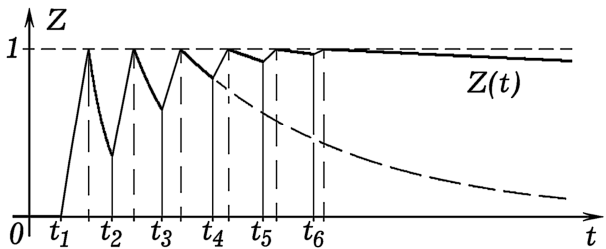

Рисунок 1. Изменение уровня знаний одного ЭУМ в результате 6 повторений

### III. Обучение в результате многократного повторения множества ЭУМ на одном уроке

Теперь промоделируем изучение $N$ ЭУМ в течение одного урока длительностью $T$. Например, ученик изучает $N$ новых слов иностранного языка (ЭУМ), которые пронумерованы от 0 до $N$. Читая текст, он в момент $t_1$ встречается со словом 2 и в течение времени $\tau_2$ переводит его, в момент $t_2$ встречается со словом 5 и в течение времени $\tau_5$ переводит его, в момент $t_3$ – со словом 1 и т.д. Когда ученик переводит $i$-ое слово первый раз ($s_i = 1$), он обращается к словарю и выписывает значение слова, второй раз — смотрит в тетрадь, третий раз — переводит по памяти и т.д., с каждым разом затрачивая меньшее время $\tau_i$. Будем считать, что эти ЭУМ случайным образом встречаются ученику и на работу с $i$-ым ЭУМ он затрачивает время $\tau_i = 1 + 2e^{-s_i/2}$ УЕВ (время решения задачи), где $s_i$ — число обращений. По мере увеличения $s_i$ происходит уменьшение коэффициента забывания $i$-ого ЭУМ по закону $\gamma_i = 0{,}002 e^{-s_i/3}$ УЕВ.

Используется программа ПР–2; она строит графики: 1) зависимости суммарных знаний $Z$ от времени; 2) среднего времени решения задачи $\tau$ по всем ЭУМ от времени; 3) среднего коэффициента забывания $\gamma$ по всем ЭУМ от времени. Получающиеся кривые при $N=10$ и $T=300$ УЕВ изображены на рис. 2. Видно, что во время обучения суммарный уровень знаний в среднем повышается, среднее время решения задачи $\tau$ снижается, стремясь к своему пределу $\tau_\infty$, средний коэффициент забывания $\gamma$ уменьшается, стремясь к нулю.

ПР–2.
```
Program Izuchen_N_EUM;
{$N+}Uses crt, graph;
Const N=13; dt=0.005; Mt=1;
Var t,tt,g,t2,SZ,SZ1,ST,Sg: single;
i,j,Gd,Gm: integer; s: array[1..N] of integer;
Z: array[1..N] of single;
BEGIN Gd:= Detect; InitGraph(Gd,Gm,'c:\bp\bgi');
Randomize; t:=-3;
Repeat t:=t+dt; tt:=tt-dt;
For i:=1 to N do If i<>j then
     Z[i]:=Z[i]-2E-3*exp(-s[i]/3)*Z[i];
If (tt<=0)and(t>50)and(t<350) then begin
j:=round(random(N*10)/10)+1;
If j>N then j:=1; Z[j]:=1; inc(s[j]);
tt:=1+2*exp(-s[j]/2); t2:=t; end;
SZ:=0; For i:=1 to N do SZ:=SZ+Z[i];  ST:=0;
For i:=1 to N do ST:=ST+(1+2*exp(-s[i]/2))/N;
Sg:=0;
For i:=1 to N do Sg:=Sg+(2E-3*exp(-s[i]/3))/N;
If t<350 then line(10+round(Mt*t),480-round
(30* SZ),10+round(Mt*t2),480-round(30*SZ1));
If t<350 then circle(10+round(Mt*t),485,1);
circle(10+round(Mt*t),480-round(8E+4*Sg),1);
circle(10+round(Mt*t),480-round(80*ST),1);
circle(10+round(Mt*t),480-round(30*SZ),1);
circle(10+round(Mt*t),480,1); SZ1:=SZ;
circle(10+round(Mt*t),200-round(100*Z[2]),1);
until (KeyPressed){or(t>700)};
{writeln(t,' Znaniya ',SZ,' zabivan ',Sg);
readkey;} CloseGraph;
END.
```

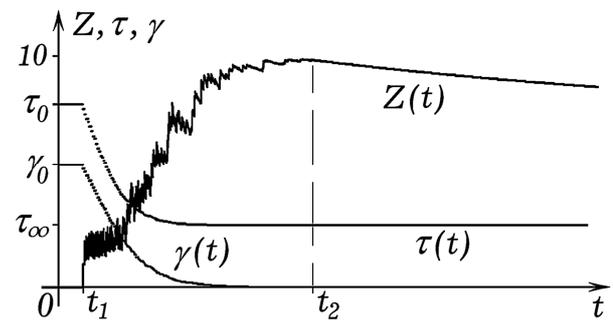

Рисунок 2. Результаты моделирования при $N = 10$.

Результаты моделирования для $N = 6$ и 10 представлены на рис. 3. Так как длительность урока одна и та же, то при увеличении количества $N$ изучаемых ЭУМ число обращений к каждому ЭУМ уменьшается, они запоминаются менее прочно, то есть средний коэффициент забывания $\gamma$ в конце обучения слишком велик. Поэтому при $N = 10$ уровень знаний $Z(t)$ после окончания обучения снижается вследствие забывания, а при $N = 6$ остается практически постоянным.

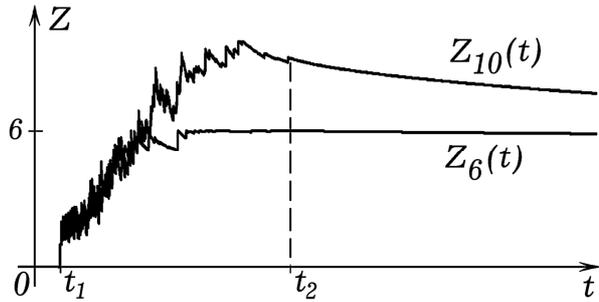

Рисунок 3. Изменение уровня знаний при $N = 6$ и 10.

При еще большем количестве изучаемых ЭУМ кривая $Z(t)$ после окончания обучения быстро убывает. На рис. 4 приведен результат моделирования при $N = 20$. Итак, анализируемая модель показывает, что количество $N$ ЭУМ, изучаемых на 1 уроке, не должно быть слишком велико. При больших $N$ знания усваиваются непрочно и потом быстро забываются.

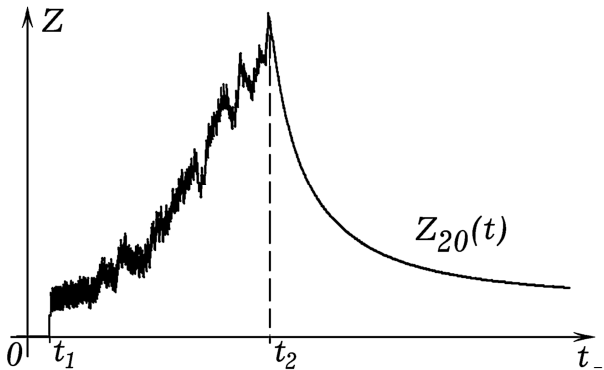

Рисунок 4. Изменение уровня знаний при $N = 20$.

Теперь изучим зависимость уровня знаний $Z$ ученика через некоторое время $t'$ после окончания урока от числа $N$ изученных ЭУМ (или скорости поступления учебной информации $v = N/T$). Изменим программу ПР–2 так, чтобы изучаемые ЭУМ следовали бы по порядку, а не случайным образом, и рассчитаем уровень знаний ученика $Z$ и средний коэффициент забывания $\gamma$ при различных $N$. В нашем случае длительность урока составляла $T = 300$ УЕВ, а контрольное время после окончания обучения $t' = 350$ УЕВ. Также вычислим показатель эффективности $K = Z/N$, равный отношению количества знаний $Z$ в момент $T + t'$ к общему числу изученных ЭУМ $N$.

Результаты моделирования приведены на рис. 5. Видно, что с ростом $N$ от 3 до 21 средний коэффициент забывания $\gamma$ растет от $2 \cdot 10^{-17}$ до $4 \cdot 10^{-5}$, то есть ЭУМ в среднем усваиваются хуже, забываются быстрее. При $N < 12$ количество знаний ученика $Z$ через время $t'$ после окончания обучения с ростом $N$ увеличивается, достигает максимума при $N = 12$, а затем при $N > 12$ уменьшается. Это объясняется влиянием двух факторов: 1) увеличением числа $N$ изученных ЭУМ; 2) уменьшением количества обращений к каждому ЭУМ в течение урока фиксированной длительности $T$ и, как следствие, ухудшением качества усвоения знаний (увеличением коэффициента забывания $\gamma$). При небольших $N$ показатель эффективности $K$ равен 1, а с ростом $N$ снижается до нуля. Итак, существует такое $N$, при котором уровень знаний $Z$ ученика через время $t'$ после окончания урока длительности $T$ будет максимальным. Чтобы обучение было эффективным, необходимо найти оптимальное значение скорости поступления учебной информации $v = N/T$.

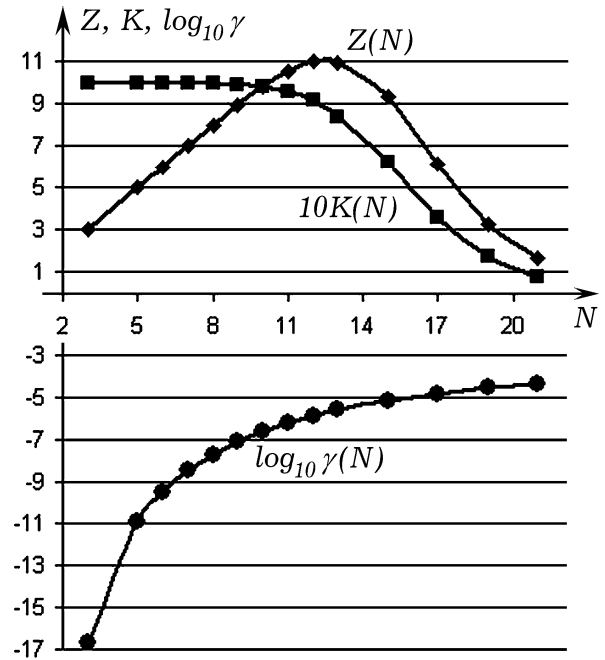

Рисунок 5. Результаты моделирования обучения на уроке при различных N.

## IV. Обучение на нескольких уроках

Теперь промоделируем изучение $N = 30$ ЭУМ в течение 3 уроков длительностью $T = 180$ УЕВ, разделенных перерывами продолжительностью $T_n = 220$ УЕВ. Во время урока ученик обращается то к одному, то к другому ЭУМ с равными вероятностями. По мере роста числа $s_i$ обращений к $i$–ому ЭУМ уменьшается затрачиваемое время $\tau_i$ и коэффициент забывания $\gamma_i$.

ПР–3.
```
program Izuch_30_EUM_3_uroka;
{$N+}Uses crt, graph;
Const N=30; dt=0.003; Mt=0.5;
Var t,g,tt,SZ: single; i,j,Gd,Gm: integer;
s: array[1..N] of integer; Z: array[1..N] of single;
BEGIN Gd:= Detect; InitGraph(Gd,Gm,'c:\bp\bgi');
Randomize; t:=-20; j:=1;
Repeat t:=t+dt; tt:=tt-dt;
If (tt<=0)and(((t>0)and(t<180))or((t>400)and
(t<580))or((t>800)and(t<980))) then begin
j:=round(random*N); If j>N then j:=N;
If j<1 then j:=1; {inc(j); If j>N then j:=1;}
Z[j]:=1; inc(s[j]); tt:=1+2*exp(-s[j]/2); end;
For i:=1 to N do Z[i]:=Z[i]-2E-3*exp(-s[i]/1.5)*Z[i];
SZ:=0; For i:=1 to N do SZ:=SZ+Z[i];
circle(20+round(Mt*t),170-round(5*j),1);
```

```
circle(20+round(Mt*t),170,1);
circle(20+round(Mt*t),500-round(10*SZ),1);
circle(20+round(Mt*t),500,1);
until KeyPressed; CloseGraph;
END.
```

Используется программа ПР–3 (среда Free Pascal), результаты имитационного моделирования приведены на рис. 5. Видно, что во время обучения номер $i$ рассматриваемого ЭУМ изменяется случайно от 1 до 30, уровень знаний ученика $Z$ при этом увеличивается. Во время перерывов происходит забывание, $Z$ уменьшается.

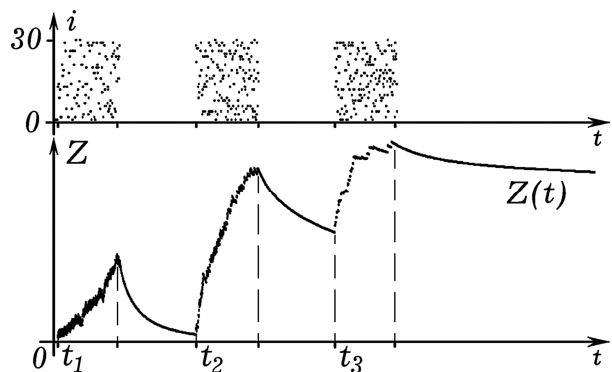

Рисунок 5. Результаты моделирования изучения 30 ЭУМ на трех уроках.

## V. Заключение

Предложенная компьютерная модель обучения, учитывает, что при увеличении числа обращений ученика к данному элементу учебного материала: 1) время его использования уменьшается, стремясь к некоторому пределу; 2) коэффициент забывания уменьшается, стремясь к нулю. Эта модель не имеет аналогов в известной автору литературе. Она позволяет промоделировать: 1) обучение в результате многократного повторения одного ЭУМ на уроке; 2) обучение в результате многократного повторения множества ЭУМ на одном уроке; 3) обучение в результате многократного повторения множества ЭУМ на нескольких уроках. При этом показано, что с ростом скорости поступления учебной информации уровень знаний ученика через время $t'$ после окончания обучения сначала увеличивается, достигает максимума, а затем уменьшается.

Следует отметить, что рассмотренные выше и подобные им [1 – 9] компьютерные модели процесса обучения дополняют качественные рассуждения, делают их более объективными, обоснованными и могут быть использованы тогда, когда проведение педагогического эксперимента требует больших затрат, неправомерно или приводит к отрицательным результатам. Изменяя последовательность изучения различных ЭУМ, длительность занятий и т.д., можно с помощью компьютерной модели найти оптимальный путь обучения в конкретном случае.

Одно из направлений использования имитационного моделирования процесса обучения связано с созданием обучающей программы, моделирующей учебный процесс в школе, которая предназначена для тренировки студентов педагогических вузов. Она должна допускать изменение параметров учеников, длительность занятий, распределения учебного материала и стратегии поведения учителя. В процессе ее работы студент, играющий роль учителя, изменяет скорость подачи учебной информации, быстро реагирует на вопросы учеников, проводит контрольные работы, ставит оценки, пытаясь добиться наибольшего уровня знаний за заданное время. После окончания "обучения" на экран выводятся графики, показывающие изменение "количества знаний учеников класса", оценки за "выполненные контрольные работы" и т.д. Кроме того, обучающая программа может проанализировать работу "учителя" (студента) и поставить ему оценку.

# Computer model of teaching with the varied coefficient of forgetting

Mayer R.V.

*Abstract*— At computer modeling of process of training it is usually supposed that all elements of a training material are forgotten with an identical speed. But in practice that knowledge which are included in educational activity of the pupil are remembered much more strongly and forgotten more slowly then knowledge which he doesn't use. For the purpose of more exact research of didactic systems is offered the model of training, in which consider that in case increasing the number of applications of this element of a learning material: 1) duration of its use by the pupil decreases; 2) the coefficient of forgetting decreases. The computer model is considered, programs in the Pascal language are submitted, results of modeling are given and analyzed.

*Keywords*— didactics, information and cybernetic approach, computer modeling of process of training.